\documentclass[twocolumn,showpacs,preprintnumbers,amsmath,amssymb,prb,aps]{revtex4-1}
\usepackage{epsfig}
\usepackage{epstopdf}
\usepackage{multirow}
\usepackage{graphicx}
\usepackage{dcolumn}
\usepackage{bm}
\usepackage{textcomp}
\usepackage{array}
\usepackage{romannum}
\usepackage{csquotes}
\usepackage{subcaption}
\usepackage{booktabs}
\usepackage{amsmath}
\usepackage{mathtools}
\usepackage{enumitem}
\usepackage{hyperref}

\usepackage{color}
\hypersetup{
    colorlinks=true,
    linkcolor=blue,
    filecolor=magenta,      
    urlcolor=blue,
    pdftitle={},
    }
    
\begin{document}
\title{\textbf    {Understanding the transport behaviour of PbSe: A combined experimental and computational study }}
\author{Isha Sihmar$^{1,}$}
\altaffiliation{Electronic mail: ishasihmar93@gmail.com}

\author{Abhishek Pandey$^{2,}$}
\altaffiliation{Electronic mail: vivekiit.19.mandi@gmail.com}

\author{Vinod Kumar Solet$^{2,}$}
\altaffiliation{Electronic mail: vsolet5@gmail.com}

\author{Neeru Chaudhary$^{1,}$}
\altaffiliation{Electronic mail: neeru369@pu.ac.in}

\author{Navdeep Goyal$^{1,}$}
\altaffiliation{Electronic mail: ngoyal@pu.ac.in}

\author{Sudhir K. Pandey$^{2,}$}
\altaffiliation{Electronic mail: sudhir@iitmandi.ac.in}

\affiliation{$^{1}$Department of Physics, Panjab University, Chandigarh - 160014, India}
\affiliation{$^{2}$School of Mechanical and Materials Engineering, Indian Institute of Technology Mandi, Kamand - 175075, India}

\date{\today}

\begin{abstract}
Lead chalcogenides are the promising thermoelectric (TE) materials having narrow band gap. The present work investigates the TE behaviour of PbSe in the temperature range 300-500 K. The transport properties of the sample have been studied using the Abinit and BoltzTrap code. The experimentally observed value of \textit{S} at 300 and 500 K is found to be $\sim$ 198 and 266 $\mu$V K$^{-1}$, respectively. The rate of increase in \emph{S} from 300 to 460 (460 to 500) K is found to be $\sim$ 0.4 (0.09). The temperature dependent electrical conductivity \textit{($\sigma$)} shows the increasing trend, with values of $\sim $ 0.35 $\times $ 10$^{3}$ and $\sim$ 0.58 $\times$ 10$^{3}$ $\Omega$$^{-1}$ m$^{-1}$ at 300 and 500 K, respectively. Further, the value of thermal conductivity \textit{($\kappa$)} at 300 (500) K is found to be 0.74 (1.07) W m$^{-1}$ K$^{-1}$. The value of \textit{$\kappa$} is found to be increasing upto 460 K and then starts decreasing. The dispersion plot indicates that PbSe is a direct band gap semiconductor with band gap value of 0.16 (0.27) eV considering spin-orbit coupling (without SOC). The partial density of states (PDOS) plot shows that Pb 6p and Se 4p states have a major contribution in the transport properties. The observed and calculated values of \textit{S} gives a good match for SOC case. The calculated \textit{$\sigma$} and electronic part of thermal conductivity (\textit{$\kappa{_e}$}) gives good match with the experimental data. The maximum power factor (PF) value of $\sim$ 4.3 $\times$ 10$^{-5}$ W/mK$^{2}$ is observed at 500 K. This work helps in understanding the TE behaviour of PbSe through a novel and insightful alliance of experimental measurements and DFT approach.

Keywords: thermoelectric properties, density functional theory, spin-orbit coupling, electronic structure, power factor.

\vspace{0.2cm}
 
\end{abstract}

\maketitle

\section{Introduction}
In the pursuit of sustainable and efficient energy sources, thermoelectric (TE) materials have emerged as promising candidate for their ability to convert waste heat into useful electrical energy. These materials work on the principle of TE effect, wherein a temperature gradient generates a voltage difference across the material, enabling direct energy conversion. The efficiency of a TE material is closely related to the material's dimensionless figure of merit(ZT)

\begin{equation} 
 ZT = \frac{S^2\sigma T} {\kappa}     \label{eq1}
 \end{equation}
 
  where \textit{S}, \textit{$\sigma$} and T stands for Seebeck coefficient, electrical conductivity and absolute temperature. The thermal conductivity (\textit{$\kappa$ = $\kappa$$_e$ + $\kappa$$_l$}) is contributed by charge carriers (\emph{e}) and lattice vibrations (\emph{l}), respectively. Further, \textit{S$^{2}$$\sigma$} represents the power factor of the material \cite{mahan1996best, pei2011convergence,pandey2023exploring}. For high \textit{ZT}, a material should have high \textit{$\sigma$}, large \textit{S} and low \textit{$\kappa$}. The transport parameters are interdependent, change in one may affect the other parameter\cite{gayner2016recent,tan2016rationally}. So, enhancing the value of \textit{ZT} is a challenging task. Currently, the low efficiency limits the technology's potential for large-scale application. The development of TE materials with desirable \textit{ZT} values is a prerequisite for advancing the genuine applications of TE technology \cite{qin2021power, sun2022strategies}.   

The most researched TE materials at low temperature are alloys based on Bi$_2$Te$_3$. Furthermore, PbTe, GeTe are the promising materials in the medium temperature ranging from $\sim$ 500-900 K and SiGe based materials are used at high temperature ($>$ 900 K)\cite{snyder2008complex}. For Bi$_2$Te$_3$ alloys, \textit{ZT} value of $\sim$ 1.1 and 1.2 is observed for p and n type doping \cite{bano2020room, zhu2021point}, respectively. For PbTe based alloys, \textit{ZT} value as high as 2.5 at 923 K and $\sim $ 1.5 at 773 K have already been reported \cite{tan2016non, liu2022high}. PbTe have narrow band gap and low thermal conductivity (\emph{$\kappa$}), which makes it useful in various applications such as optical computing, photovoltaic cells, solar cells and TE devices\cite{khokhlov2021lead,ravich2013semiconducting}. The scarcity of Te in earth-crust, makes it important to look for Te free chalcogenides. Se is more abundant and cheaper than Te.     
PbSe has been recognised as a narrow band gap semiconductor with a direct band gap energy of 0.27 eV \cite{achimovivcova2009characterization}. PbSe has a high melting point of 1340 K as compared to PbTe for 1190 K, which makes it more chemically stable. This has drawn the attention of researchers towards PbSe as an alternate for PbTe. 
 Researchers have been working to study the properties of PbSe \cite{alekseeva1996thermoelectric, ibrahim2008electrical}. But, for undoped PbSe, \textit{ZT} value is still less than one which is not significant to make PbSe efficient TE material. Hence, to understand TE properties of the material a combined experimental and computational study is required. 


TE properties of PbSe have been studied both experimentally and computationally. Jingyang  \emph{et~al.}\cite{du2020rapid} synthesized single phase PbSe via mechanical alloying and high-pressure sintering, and observed a \textit{ZT} value $\sim $ 0.57 at 600 K, similarly \textit{ZT} value of 0.55 was reported at $\sim $ 475 K by Chen \emph{et~al.} for high temeprature high pressure (HTHP) sintered\cite{chen2018effects} sample. Furthermore, Wang \emph{et~al.} studied TE properties of undoped and doped PbSe prepared via melt-quench technique and reported a high \textit{ZT} value of 0.58 at 465 K and 1 at $\sim$ 770 K for undoped and Ag doped PbSe, respectively\cite{wang2011exploring}. 
Meanwhile Tang \emph{et~al.} explored the computational approach to study the properties of p-type PbSe monolayer within the framework of Density functional theory (DFT) and estimated a \emph{ZT} value of $\sim$ 1.3 at 900 K\cite{tang2022honeycomb}. In contrast, Parker \emph{et~al.} proposed that heavily doped PbSe may show \emph{ZT} value as high as 2 at $\sim$ 1000 K\cite{parker2010high}. 
The TE properties of PbSe have been explored individually, either experimentally or computationally. However, there is a significant difference between the experimental and computationally observed values of transport parameters. Gayner \emph{et~al.}\cite{gayner2016boost} and Wang \emph{et~al.}\cite{wang2011exploring} reported that the experimental value of \emph{S} at 300 K is $\sim$ 248 $\mu$V K$^{-1}$ and $\sim$ 182 $\mu$V K$^{-1}$, respectively. Whereas, Khan \emph{et~al.} \cite{khan2016thermoelectric} reported the computationally observed value of \emph{S} at 300 K to be $\sim$ 316.5 $\mu$V K$^{-1}$. Here, one can observe that the reported experimental and computational values are not in good agreement. 
Many key factors like choice of exchange-correlation functional, pseudo potentials, convergence parameters and SOC play a significant role in the computational study of TE materials.
By paying close attenion to these parameters there should be better match between the experimental and computational results. Therefore, a systematic experimental investigation, supported by the computational study is required to understand the properties of PbSe.  

This research paper aims to present a comprehensive investigation into the TE properties of PbSe through a novel and insightful alliance of experimental measurements and computational simulations. By examining electrical conductivity (\emph{$\sigma$}) and seebeck coefficient (\emph{S}) and thermal conductivity (\emph{$\kappa$}), we aim to unravel the intricate interplay of charge carriers, electronic band structures and density of states (DOS) that govern TE performance of PbSe. The experimental methodology involves the synthesis of PbSe sample and subsequent characterization using the in-house instruments\cite{sk2022instrument,singh2018fabrication} along with thorough study of structural properties. The value of \emph{S} at 300 K is observed to be $\sim$ 198 $\mu$V K$^{-1}$ and it increases with temperature to $\sim$ 266 $\mu$V K$^{-1}$ at 500 K. Positive value of \emph{S} in the whole temperature range indicates the p type behaviour of PbSe. Value of \emph{$\sigma$} and \emph{$\kappa$} increases as the temperature increases, with \emph{$\sigma$} (\emph{$\kappa$}) value at 300  and 500 K is observed to be $\sim $ 0.35 $\times $ 10$^{3}$ (0.74) and $\sim$ 0.58 $\times$ 10$^{3}$ $\Omega$$^{-1}$ m$^{-1}$ (1.07 W m$^{-1}$ K$^{-1}$), respectively. The observed power factor (PF) values are in the range of 1.39 $\times$ 10$^{-5}$ to 4.34 $\times$ 10$^{-5}$ W/mK$^{2}$. To comprehend the experimental behaviour, electronic structure calculations have been performed using DFT. Transport properties have been calculated using BoltzTraP code. Desity of states (DOS) plot shows that p states of both the elements have major contribution in the transport properties. 
The calculated \emph{S} is found to be in good agreement with the experimental values. The calculated \emph{$\sigma$} and electronic part of thermal conductivity (\textit{$\kappa{_e}$}) gives good match with the experimental data.
The maximum observed PF has been found to be $\sim$ 192 $\times$ 10$^{-4}$ W/mK$^{2}$ at 900 K.  
 The results of this combined experimental and computational study are expected to contribute significantly to the understanding of TE behaviour of PbSe system.  
\section{Experimental and Computational details}
PbSe sample was prepared employing melt quench technique \cite{duwez1960continuous,anantharaman1971decade,tan2016non}. The high purity Pb (99.95\%), Se (99.99\%) were stoichiometrically weighed and poured into the quartz ampoules (outer and inner diameter $\sim$ 1 cm and 0.7 cm, respectively). The ampoules were vacuum sealed under pressure of 10$^{-4}$ to 10$^{-5}$  mbar and then placed in the furnace at temperature 1073 K, which was maintained for 24 hours. Afterwards, the ampoules were quenched using ice cold water. The ingots thus obtained were grinded in mortar and pestle to get fine powder. XRD on the sample was performed via a Rigaku X-ray diffractometer using Cu-K$_\alpha$ radiation from the range of 10$^\circ$-80$^\circ$ with scan speed of 2$^\circ$/min. To further solidify information obtained from qualitative analysis of XRD data, Reitveld refinement was done using X'pert Highscore Plus software \cite{degen2014highscore}. The circular and rectangular pellets of dimensions (10 mm $\times$ 1 mm) and (6 mm $\times$ 4 mm $\times$ 1 mm) were prepared under 40 kN m$^{-2}$ pressure for \emph{S}, \emph{$\sigma$} and \emph{$\kappa$} measurements, respectively.  
These measurements were performed in the range of 300-500 K using in-house setups \cite{sk2022instrument,singh2018fabrication}. 

For better understanding of the experimental results, theoretical calculations were adopted to study the dispersion curve, DOS, electronic transport properties. For the electronic structure and DOS calculations the projected augmented wave (PAW) method \cite{blochl1994projector} has been used in the ABINIT software \cite{gonze2002first}. The Perdew, Burke, and Ernzerhof (PBE) functional with Generalised Gradient approximation (GGA) \cite{perdew1996generalized} were used as exchange-correlation (XC) functional in the self consistent calculations. The lattice parameters a = b = c = 6.10 A$^\circ$ obtained from the Reitveld refinement were used for the calculations. The transport properties were calculated using BoltzTraP code \cite{madsen2006boltztrap} based on semi-classical Boltzmann transport theory. In order to get converged transport properties a dense k-point mesh of 50 $\times$ 50 $\times$ 50 is used. 

\section{Results and Discussion}
\subsection{XRD analysis}


\begin{figure}[htb!]\centering
\includegraphics[width=0.9\linewidth, height=12cm]{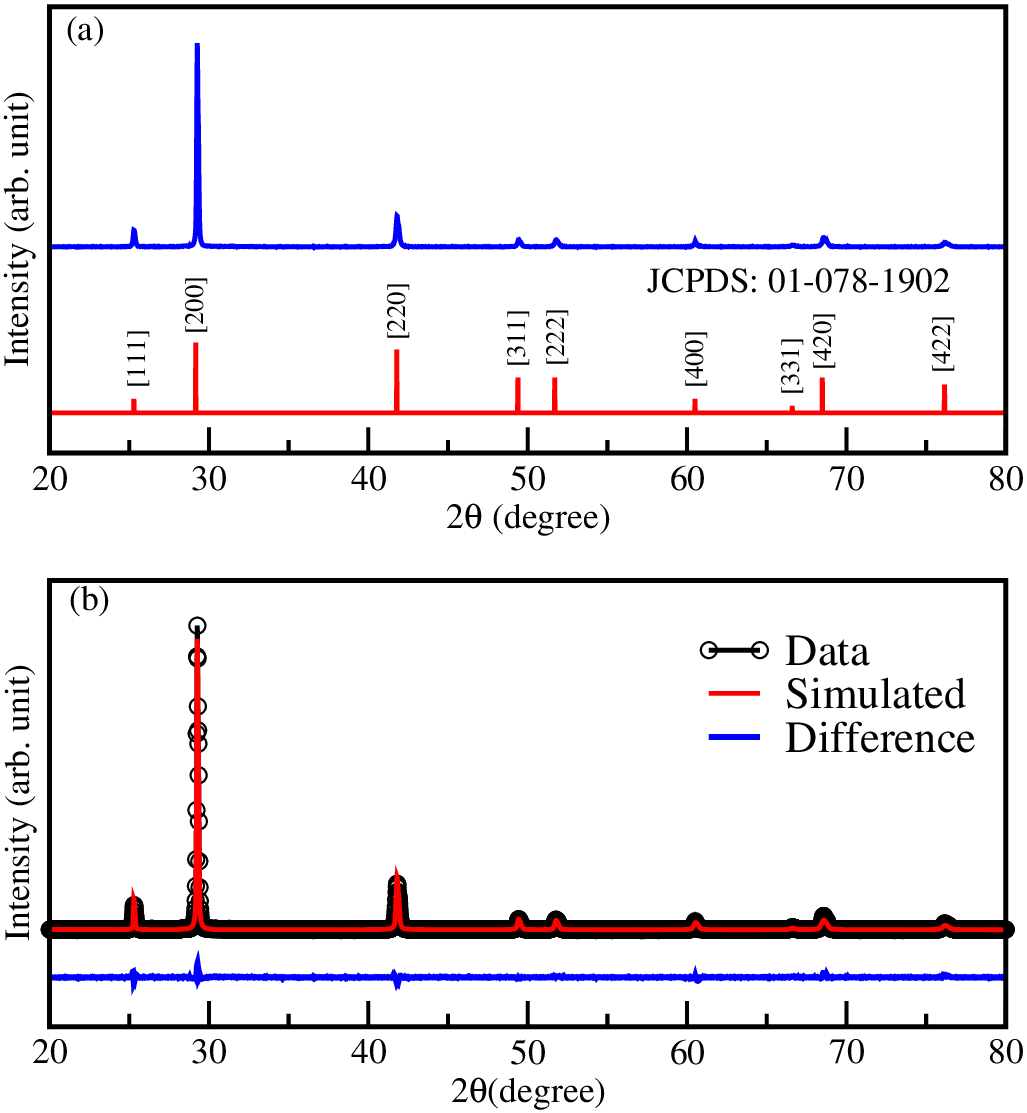}
\caption{(a) X-ray Diffraction of PbSe. (b) Reitveld refinement plot for PbSe} \label{fig1} 
\end{figure}

The room temperature XRD pattern of synthesized PbSe is shown in Fig.\ref{fig1}(a). JCPDS: 01-078-1902 is used to compare the XRD data and pure phase formation of the system has been expected as there is no extra peak indicating the impurity. The (hkl) values corresponding to each peak is labelled in the Fig. with reference to JCPDS used. For phase identification, the XRD pattern was further studied by Reitveld refinement using X’pert high score plus software\cite{degen2014highscore}. Fig.\ref{fig1}(b) shows the Reitveld refinement of the XRD data. 
From refinement it was confirmed that the synthesized sample was in single phase. The lattice parameters obtained from refinement are a = b = c = 6.10 A$^\circ$ and were found to be in good aggrement with the previously reported literature \cite{nikolic1969solid}.


\subsection{\label{sec:level2}Experimental Transport Properties}



\begin{figure}[htb!]\centering
\includegraphics[width=0.87\linewidth, height=18cm]{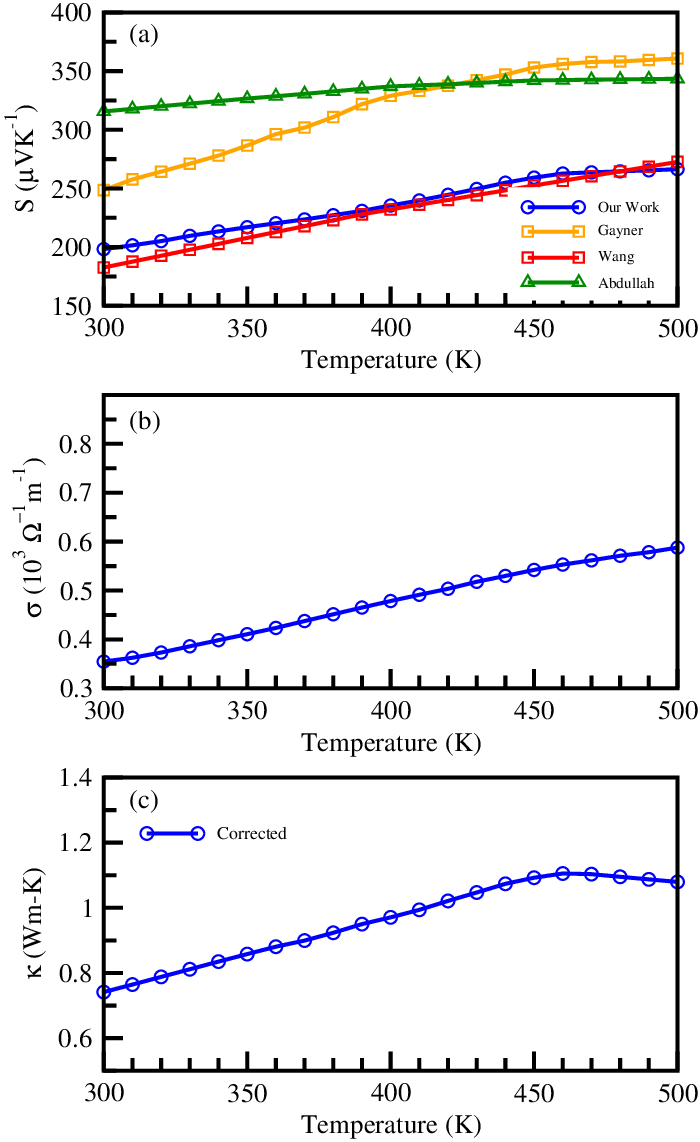}
\caption{Temperature dependence of (a) {Seebeck Coefficient}  (b) {Electrical Conductivity}  and (c) {Thermal Conductivity} }\label{fig2}
\end{figure}

Fig.\ref{fig2}(a) is representing the experimentally observed \emph{S} with varying temperature from 300 to 500 K. The positive value of \emph{S} in the entire observed temperature range indicates the p-type behaviour of PbSe, which indicates that holes are the dominating charge carriers in the compound. In this work, the magnitude of \emph{S} increases with the rise in temperature. The measured value of \emph{S} at 300 and 500 K is observed to be $\sim$ 198 and $\sim$ 266 $\mu$V K$^{-1}$, respectively. Furthermore, the rate in the increase of \emph{S}  from 300 to 460 (460 to 500) K is found to be $\sim$ 0.4 (0.09). The observed \emph{S} matches well with the previous reported work of Wang \emph{et~al.} \cite{wang2011exploring} (this work) having values of $\sim$ 182 (198) and 272 (266) $\mu$V K$^{-1}$ at 300 and 500 K, respectively. But, no change in the increment rate of \emph{S} around 460 K has been observed in the Wang \emph{et~al.} work.
 Although, difference in the magnitude of \emph{S} in Gayner \emph{et~al.}\cite{gayner2016boost} and this work have been observed. Both shows the similar trend in the observed temperature range of 300-500 K. In Gayner \emph{et~al.} work the increment rate of \emph{S} from 300 to 450 (450-500) K is found to be  0.7 (0.14). Here one can observe that the ratio between the increment rate of \emph{S} from 300 to 450 K with respect to the rate of increment from 450 to 500 K is found to be  $\sim$ 5. Furthermore, in our work this ratio has been found to be 4.4, which is similar to Gayner \emph{et~al.} work. The further understanding of \emph{S} have been explored in the computational part.

Fig.\ref{fig2}(b) shows the temperature dependence of electrical conductivity (\emph{$\sigma$}). The values of \emph{$\sigma$} at 300 and 500 K are found to be $\sim$ 0.35 $\times$ 10$^{3}$ and $\sim$ 0.58 $\times$ 10$^{3}$ $\Omega$$^{-1}$ m$^{-1}$, respectively. The value of \emph{$\sigma$} is found to increase linearly with temeperature. The increase in \emph{$\sigma$} with temperature is consistent with the earlier reported work, Fan \emph{et~al.} \cite{fan2014high} (this work) reported the value of \emph{$\sigma$} at 300 and 500 K to be $\sim$ 0.71 $\times$ 10$^{3}$ (0.35 $\times$ 10$^{3}$) and $\sim$ 1.43 $\times$ 10$^{3}$ (0.58 $\times$ 10$^{3}$) $\Omega$$^{-1}$ m$^{-1}$, respectively. 
For understanding this behaviour, consider the relation between \emph{$\sigma$}, relaxation time ($\tau$), effective mass (m$^{*}$) and n given by

\begin{equation}
\sigma  = \frac{ne^2 \tau }{m^*}   \label{eq3}
\end{equation}

The increase in lattice vibrations with temperature leads to more frequent collisions which tends to reduce $\tau$ and hence decrease in \emph{$\sigma$}. It is well known that in semiconductors the band gap decreases with the rise in temperature.  
From eq.(\ref{eq3}) one can say that the increase in conductivity can be due to the rise in number of charge carriers (electrons) available for conduction. Therefore, in contrast to the effects of larger thermal vibrations, which typically decreases $\tau$ and hence \emph{$\sigma$}, the carrier concentration is expected to be dominant. 


Temperature dependence of corrected \emph{$\kappa$} by using Loeb's eq.\cite{francl1954thermal} for PbSe in the temperature range 300-500 K is shown in Fig.\ref{fig2}(c). 

\begin{equation}
\kappa  = \kappa_0 (1-\phi)   \label{eqA}
\end{equation}
 
where \emph{$\kappa$} is the observed value of thermal conductivity, \emph{$\kappa$}$_0$ is the value of thermal conductivity for material with zero porosity and $\phi$ is the volume pore fraction.
The value of \emph{$\kappa$} at 300 K is 0.74 W m$^{-1}$ K$^{-1}$. In the temperature range 300-460 K the value of \emph{$\kappa$} increases, and is observed to be $\sim$ 1.10 W m$^{-1}$ K$^{-1}$ at 460 K. Further, with the rise in temperature a slight decrease in \emph{$\kappa$} is observed and the value becomes $\sim$ 1.07 W m$^{-1}$ K$^{-1}$ at 500 K. This increasing\cite{khan2016thermoelectric} and decreasing\cite{wang2011exploring,gayner2016boost} trend in the value of \emph{$\kappa$} have been observed in the previous reported data. 
The increasing \emph{$\kappa$} with temperature has been reported by Khan \emph{et~al.} \cite{khan2016thermoelectric} (this work), value of \emph{$\kappa$} at 300 and 500 K was observed to be $\sim$ 0.27 (0.74) and $\sim$ 1.96 (1.10) W m$^{-1}$ K$^{-1}$, respectively. The total thermal conductivity consists of contribution from both the lattice thermal conductivity (\emph{$\kappa_L$}) and elctronic thermal conductiviy (\emph{$\kappa_e$}) : \emph{$\kappa$} = \emph{$\kappa_L$} + \emph{$\kappa_e$}. Further understanding of \emph{$\kappa$} have been discussed in the next sections.

\subsection{\label{sec:level3}Electronic Structure Calculations} 

For better understanding of experimentally observed TE properties, electronic structure and transport calculations have been performed. The lattice and structural parameters obtained from Reitveld refinement have been used for the calculations. Since, Pb is a heavy element, therefore SOC is expected to play a significant role in the electronic structure of the compound\cite{khan2016thermoelectric}.
SOC splits the bands, which alters the density of states (DOS) and hence, can effect the transport properties. Therefore, it has been taken into account while calculating the dispersion curve. 
Fig.\ref{Fig.3} shows dispersion curve for PbSe with and without SOC along the high symmetric points $\varGamma -\emph{X}-\emph{W}-\emph{L}-\varGamma $ within the first brillouin zone. In plot, Fermi level (E$_F$) is set at the centre of the band gap. The black and red lines represents bands without and with SOC, respectively. It can be observed from both SOC and non-SOC dispersion curve that valence band maximum (VBM) and conduction band minimum (CBM) lies at high symmetric \emph{$L$}-point. This shows that PbSe is a direct band gap semiconductor, which is in accordance with the previously reported literature\cite{albanesi2000electronic,parker2010high}. The estimated value of band gap with and without SOC is found to be $\sim$ 0.16 eV and 0.27 eV respectively, indicating that PbSe is a narrow band-gap semiconductor. Here, for PbSe the inclusion of SOC decreases the band gap. \\\\\\\\

 \begin{figure}[htb!]\centering
 \label{Fig.3}
\includegraphics[width=0.85\linewidth, height=6.2cm]{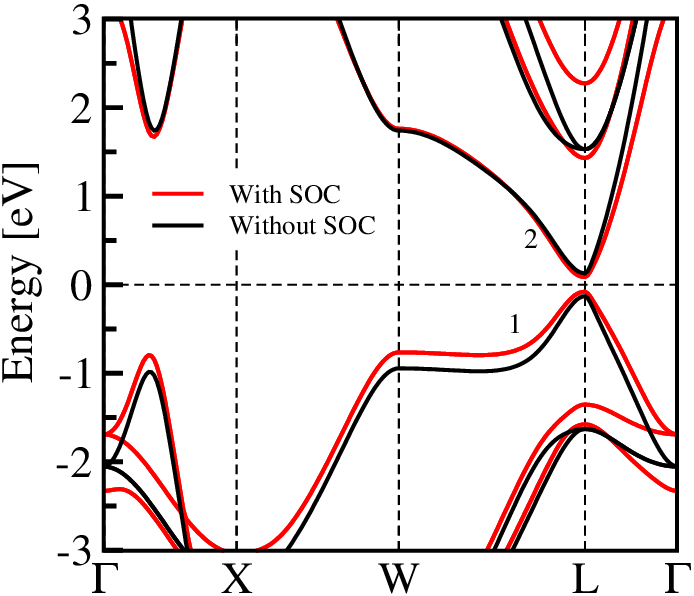}
\caption{\label{Fig.3}\small{The band structure along high-symmetric directions}}
\end{figure}

In order to investigate more about the transport properties, m$^{*}$ of the charge carriers have been evaluated. The m$^{*}$ value depicts how the energy of a charge carrier changes with respect to its momentum and is determined by the shape of energy band. The dependence of m$^{*}$ on the curvature of band can be explained by the relation m$^{*}$= $\hbar$/$(\partial$$^{2}$E/$\partial$k$^{2}$), where denominator represents the curvature of band \cite{ashcroft2011solid}. It can be stated that for the flat bands m$^{*}$ value will be greater as compared to the bands with more curvature. From Fig.\ref{Fig.3}, it is clear that bands 1 and 2 have remarkable contribution to the transport properties. Therefore, m$^{*}$ value for the bands 1 and 2 in the directions \emph{L}-\emph{L$\varGamma$}  and \emph{L}-\emph{LW} have been calculated.

Table \ref{tab1} shows the m$^{*}$ values of charge carriers in high symmetric directions. 
The calculated values of m$^{*}$ for holes is slightly greater than that of electrons. So positive value of seebeck coefficient is expected which is also observed in our case. \\\\\\\\\\
 

\begin{figure}[htb!]\centering  
\label{Fig.4}
\includegraphics[width=0.88\linewidth, height=6.4cm]{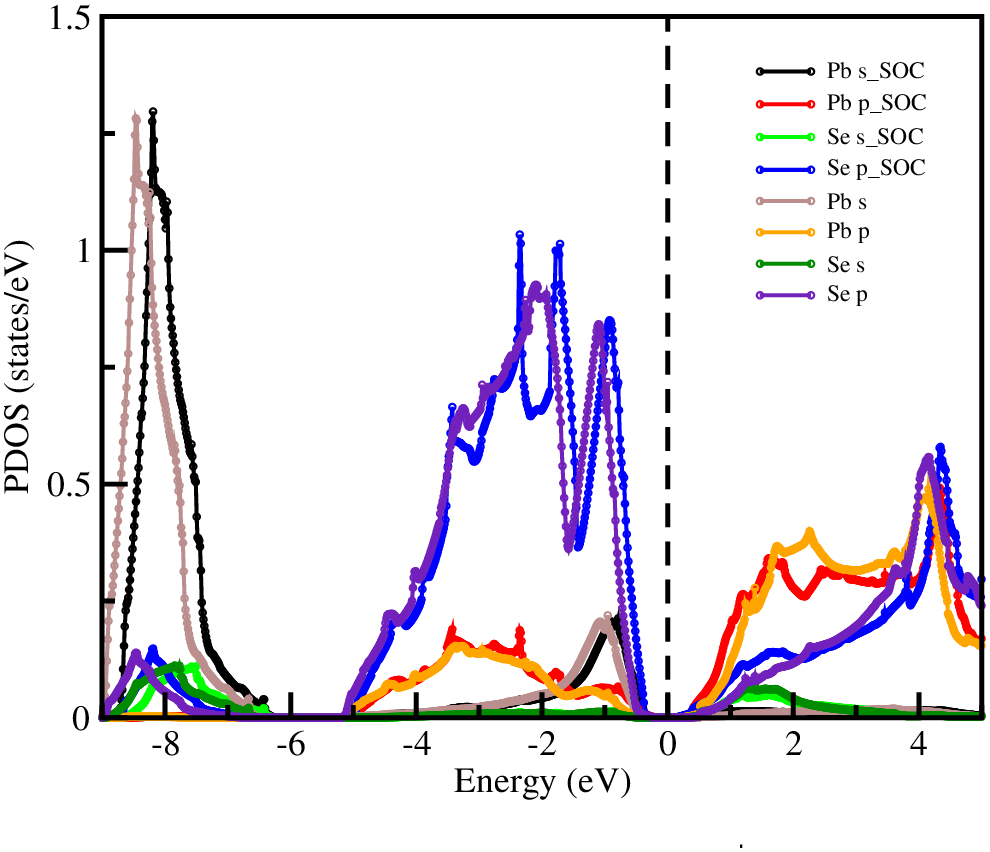}
\caption{Partial density of states with and without SOC}\label{Fig.4}
\end{figure}

\begin{table}[htb!]\centering
\caption{The effective mass of holes and electrons
(at \emph{L} - point) along high symmetric direction.}\label{tab1}
\begin{ruledtabular}
\begin{tabular}{lccc}
\textrm{\textbf{K-points}}&
\textrm{\textbf{VB}}&
\textrm{\textbf{CB}}\\
      
\colrule
  \emph{L}-\emph{L$\varGamma$}   & 0.057    & 0.040     \\
  \emph{L}-\emph{LW}               & 0.101    & 0.094     \\
\end{tabular}
\end{ruledtabular}
\end{table}

In order to observe the contribution of different atomic orbitals to the transport properties, the partial density of states (PDOS) have been calculated. Fig.\ref{Fig.4} shows the PDOS for PbSe. In the plot, E$_F$ is set at the centre of band gap. From the Fig., it can be seen that electronic states in valence band region ($\sim$ -5 eV to $\sim$ -0.15 eV) are mainly from the contribution of Se 4p state and negligible states of Pb 6p and 6s. In conduction band the electronic states is contributed from Pb 6p and a small part of Se 4p states. Near the Fermi level, there is very small contribution in the electronic states from Se 4s and Pb 6s states. This suggests that Pb 6p and Se 4p states have a major contribution on the transport properties of PbSe. The band gap along with the features of the electronic dispersion curve in the valence and conduction band regions govern the \emph{S}, \emph{$\sigma$} and \emph{$\kappa$} of a TE material. Using the dispersion curve with SOC, we attempt to explain the observed experimental behaviour of PbSe via transport calculations under Boltzmann transport theory.



\subsection{\label{sec:level4}Calculated Transport Properties}

In this section temperature dependence of seebeck coefficient (\emph{S}), electrical conductivity (\emph{$\sigma$}) and electronic thermal conductivity (\emph{$\kappa{_e}$}) are discussed. The temperature dependent transport properties are calculated using BoltzTraP code\cite{madsen2006boltztrap}, which is based on semi-classical Boltzmann transport theory. Fig.\ref{Fig.5} shows the variation of \emph{S} with $\mu$ for different values of temperature for the SOC case. The black dotted line at the middle of the curve ($\mu$ = 0 meV) shows the Fermi level. For $\mu$ = 0 meV, at 300 K the magnitude of \emph{S} have been found to be $\sim $ 122 $\mu$V K$^{-1}$. 
Positive value of \emph{S} can be understood on the basis of m$^{*}$. The eq.\ref{eq2} shows relation between m$^{*}$, carrier density (n) and absolute temperature (T). 

\begin{equation} 
 S= \frac{8\pi^2 k_b ^2}{3eh^2} (\frac{\pi}{3n}) ^{2/3}   m^*T  \label{eq2}
\end{equation} 

It is clear that there is direct dependence of \emph{S} on m$^{*}$ and the positive or negative sign of \emph{S} depends on m$^{*}$. Table \ref{tab1} shows that the calculated m$^{*}$ for holes is greater than that of electrons. Therefore, holes are the dominating carriers in the transport properties and we get positive value of \emph{S}. But this value is less than our experimentally observed value of 198 $\mu$V K$^{-1}$. We looked for the value of $\mu$ at 300 K for which value of \emph{S} matches well with our experimental value of \emph{S}. It has been found that for $\mu$ $\approx$ -121 meV, the calculated and experimental values of \emph{S} gives a good match. Similarly, for the non-SOC case, the best match have been found at $\mu$ $\sim$ -154 meV. These values of $\mu$ were further used to calculate the value of \emph{S} in the temperature range 300-500 K. Fig.\ref{Fig.6}(a) shows the comparison of experimental and calculated values. Blue line represents the experimental data. Black and red lines represents the calculated values without and with considering the SOC effect. It can be easily seen that consideration of SOC for higher atomic number elements gives good agreement with the experimental data. At low temperature, calculated values gives a good match with experimental data but as temperature increases some difference is observed. At 500 K there is a difference of $\sim$ 15 and $\sim$ 35 $\mu$V K$^{-1}$ with and without SOC, respectively. It can be because BoltzTrap use default value of $\tau$ (10 $^{-14}$ sec), which is constant at every temperature. 
At high temperature the value of $\tau$ changes because of various scattering mechanisms\cite{durczewski2000nontrivial,ahmad2010energy}. 
 Therefore, the better match between experimental and computational \emph{S} can be observed after considering temperature dependent $\tau$.  \\\\



\begin{figure}[htb!]\centering
\label{Fig.5} 
\includegraphics[width=0.8\linewidth, height=6.2cm]{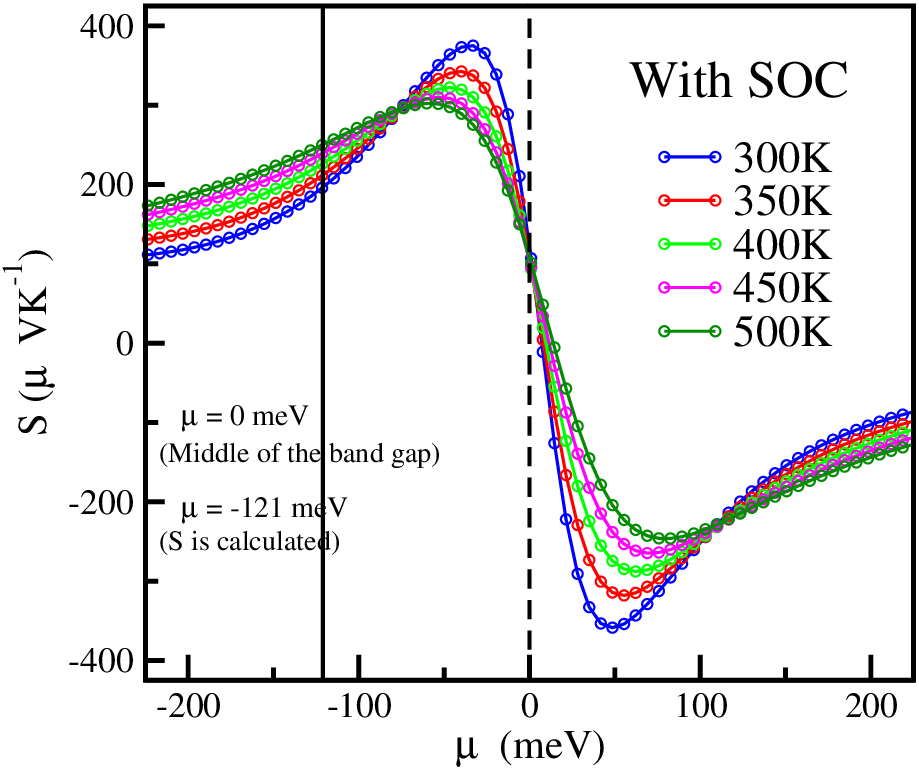}
\caption{\label{Fig.5}\small{Variation of Seebeck coefficient with Chemical Potential at different temperatures considering SOC}} 
\end{figure}

\begin{figure}[htb!]\centering
\label{Fig.6}
\includegraphics[width=0.87\linewidth,  height=18 cm]{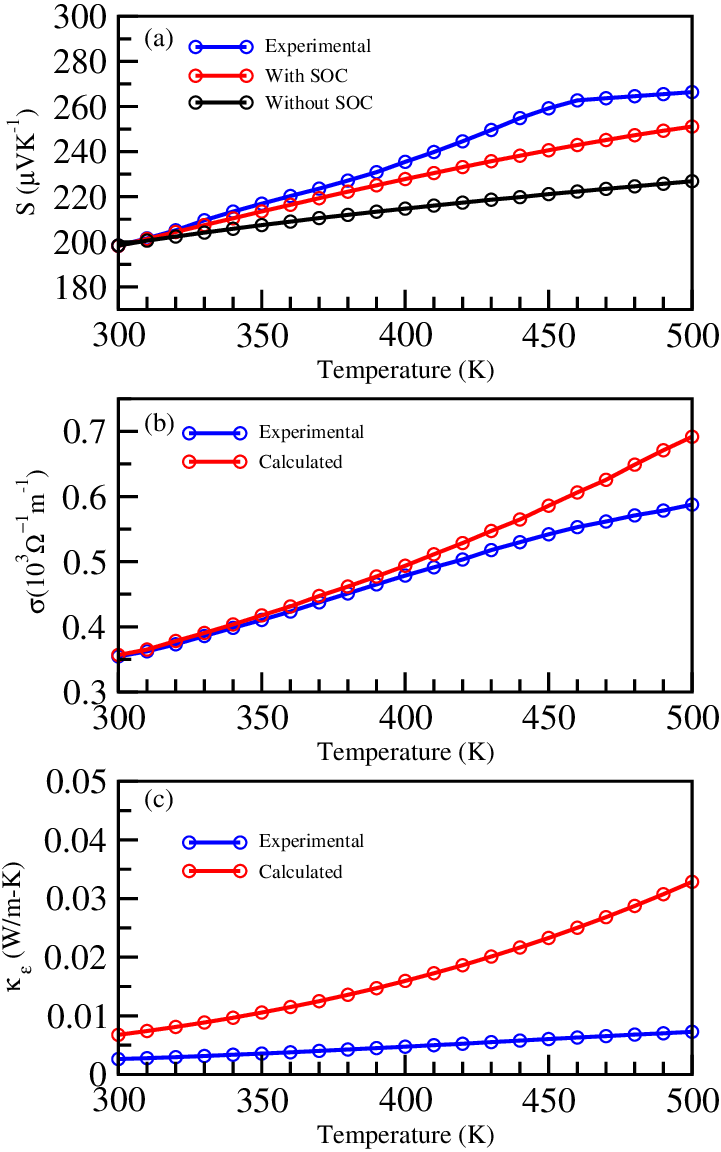}
\caption{Variation of (a) Seebeck coefficient with temperature (b) Conductivity with temperature  (c) Electronic part of thermal conductivity (Comparison of experimental and calculated data)}\label{Fig.6} 
\end{figure}
 
Fig.\ref{Fig.6}(b) shows the comparison of experimental and calculated values of \emph{$\sigma$} with varying temperature. The values of \emph{$\sigma/\tau$} in the temperature range of 300-500 K has been calculated using BoltzTrap code. The magnitude of \emph{$\tau$} used in the BoltzTraP code is $10^{-14}$ seconds. 
Initially we used this default value of \emph{$\tau$} to find \emph{$\sigma$}. However, the values obtained were not in good agreement with experimental data. 
In order to get the best match of computational \emph{$\sigma$} with the experimentally obtained \emph{$\sigma$} at 300 K, we used 1.57 $\times$ 10$^{-16}$ seconds as the value of \emph{$\tau$}. Since, the values of computational \emph{$\sigma$} is found to give reasonably good match with the experimental observed values. Therefore, the value of \emph{$\tau$} for the compound is expected to be of the order of 10$^{-16}$ seconds. 
The low value of \emph{$\tau$} could be because of the porosity of the material. Porosity increases voids within the material structure, these voids act as additional scattering centers and hence, increases the sacttering rate for the carriers as they travel through the material. This increased scattering results in more frequent interruptions in the motion of charge carriers, reducing their average time between the collisions. As a result the decreased \emph{$\tau$} reduces the overall value of \emph{$\sigma$} for PbSe.
 At high temparature there is deviation between experimental and calculated values, this maybe because of the temperature independent \emph{$\tau$}.

Comparison of experimental and calculated electronic part of thermal conductivity (\emph{$\kappa{_e}$}) is shown in Fig.\ref{Fig.6}(c). Experimental \emph{$\kappa{_e}$} is calculated using the Wiedemann-Franz Law: \emph{$\kappa{_e}$} = L${_0}$\emph{$\sigma$}T, where L${_0}$ is a constant (2.45$\times$10$^{-8}$ W$\Omega$K$^{-2}$), \emph{$\sigma$} and T values have been taken from Fig.\ref{fig2}(b).The computatioanl \emph{$\kappa{_e}$/$\tau$} is calculated under semi-classical Boltzmann theory. In order to get \emph{$\kappa{_e}$} the above proposed value of \emph{$\tau$} has been used. With increase in temperature there is only a slight increase in the value of \emph{$\kappa{_e}$}. In Fig.\ref{Fig.6}(c), the experimental (computational) value of \emph{$\kappa{_e}$} at 300, 400 and 500 K is $\sim$ 2.6 $\times$ 10$^{-3}$ (6.7 $\times$ 10$^{-3}$), 4.7 $\times$ 10$^{-3}$ (1.6 $\times$ 10$^{-2}$) and 7.2 $\times$ 10$^{-3}$ (3.3 $\times$ 10$^{-2}$) Wm$^{-1}$K$^{-1}$, respectively. Here one can observe that, with the increase of temperature the deviation between experimental and calculated values has increased. It maybe because of the temperature independent value of \emph{$\tau$} and L${_0}$. 

Typically, the value of \emph{$\tau$} decreases with increase in the temperature\cite{ahmad2010energy}. Considering the above factors, one can find the better agreement between experimental and calculated values. The experimental values of \emph{$\kappa{_{total}}$} (\emph{$\kappa{_e}$}) at 300, 400 and 500 K is 0.57 (2.6 $\times$ 10$^{-3}$), 0.74 (4.7 $\times$ 10$^{-3}$) and 0.83 (7.2 $\times$ 10$^{-3}$) Wm$^{-1}$K$^{-1}$, respectively. Therefore, it can be concluded that in the whole temperature range major contribution in total \emph{$\kappa$} is because of the lattice part (\emph{$\kappa{_l}$}). Unlike \emph{S} and \emph{$\kappa$}, no change in slope is observed near 460 K for \emph{$\kappa{_e}$}. So, the change in rate of increment in \emph{S} and \emph{$\kappa$} near 460 K can be due to the lattice part. \\\\\\
\begin{figure}[htb!]\centering
\label{Fig.7}
\includegraphics[width=0.84\linewidth, height=6.2cm]{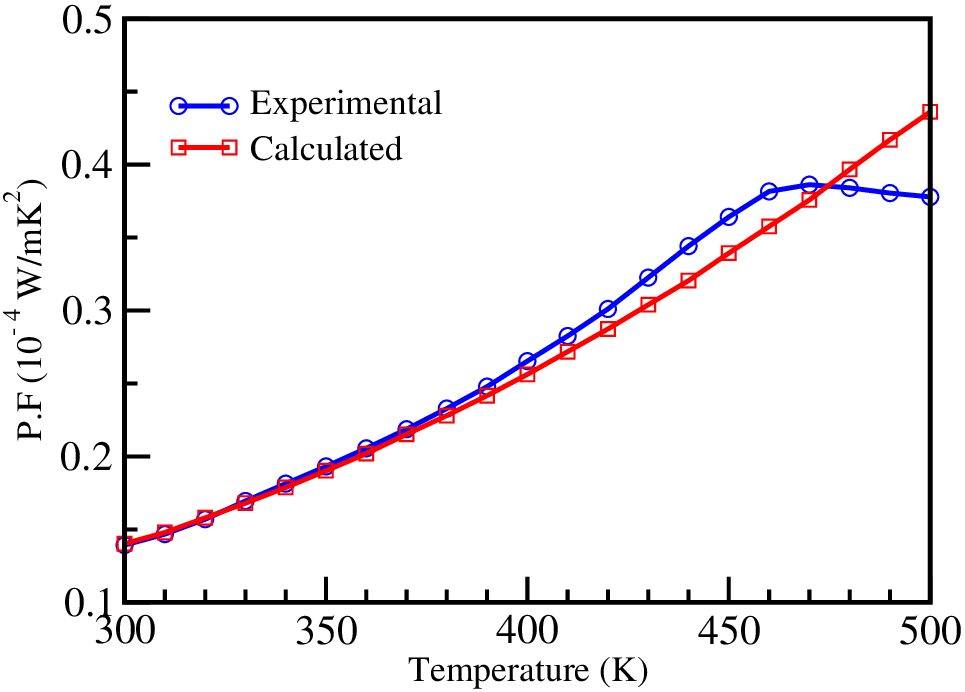} 
\caption{\label{Fig.7}\small{Variation of Power Factor with Temperature}} 
\end{figure} 

As a way to check the competence of PbSe to be used as a TE material we have evaluated the power factor (PF) values in the temperature range 300-500 K. The comparison of experimental and calculated values of PF is shown in Fig.\ref{Fig.7}. As temperature increases PF values also increases. Following the trend of \emph{S}, at low temperature PF values give a good match but as temperature increases slight deviation is observed. At 460 K the experimental value of PF starts decreasing and further after 475 K it becomes less than the calculated value. The decreasing trend in experimental PF after 460 K matches reasonably well with the previously reported data\cite{zhang2012study}. In our work PF values are found to be in the range of 1.4 $\times$ 10$^{-5}$ to 4.3 $\times$ 10$^{-5}$ W/mK$^{2}$. 


Fig.\ref{Fig.8} represents the variation in PF with $\mu$ at different temperatures, using the default value of $\tau$ implemented in BoltzTrap code. The black dotted line at the middle represents the fermi level. Positive and negative values of $\mu$ represents the n and p-type doping, respectively. Below the Fermi level, PF curve have two maxima for each value of temperature. For the first maxima below the Fermi level, maximum PF values for 300 and 900 K are $\sim$ 51 $\times$ 10$^{-4}$ W/mK$^{2}$ and $\sim$ 191 $\times$ 10$^{-4}$ W/mK$^{2}$, corresponding to $\mu$ value of $\sim$ -332 meV. For the second maxima, the maximum value of PF for 300 K is $\sim$ 64 $\times$ 10$^{-4}$ W/mK$^{2}$ at $\sim$ -766 meV. As temperature increases upto 900 K, the value of $\mu$ corresponding to maximum PF shifts towards E$_F$ and reaches upto $\sim$ -703 meV corresponding to PF value of $\sim$ 192 $\times$ 10$^{-4}$ W/mK$^{2}$ for 900 K. Above Fermi level, the value of $\mu$ corresponding to maximum PF shifts towards E$_F$ as temperature increases from 300 to 900 K. This suggests that p-type doping can increase the PF for PbSe. Consequently, it is possible to improve the TE behaviour.  \\\\\\

\begin{figure}[htb!]\centering
\label{Fig.8}
\includegraphics[width=0.8\linewidth, height=6.2cm]{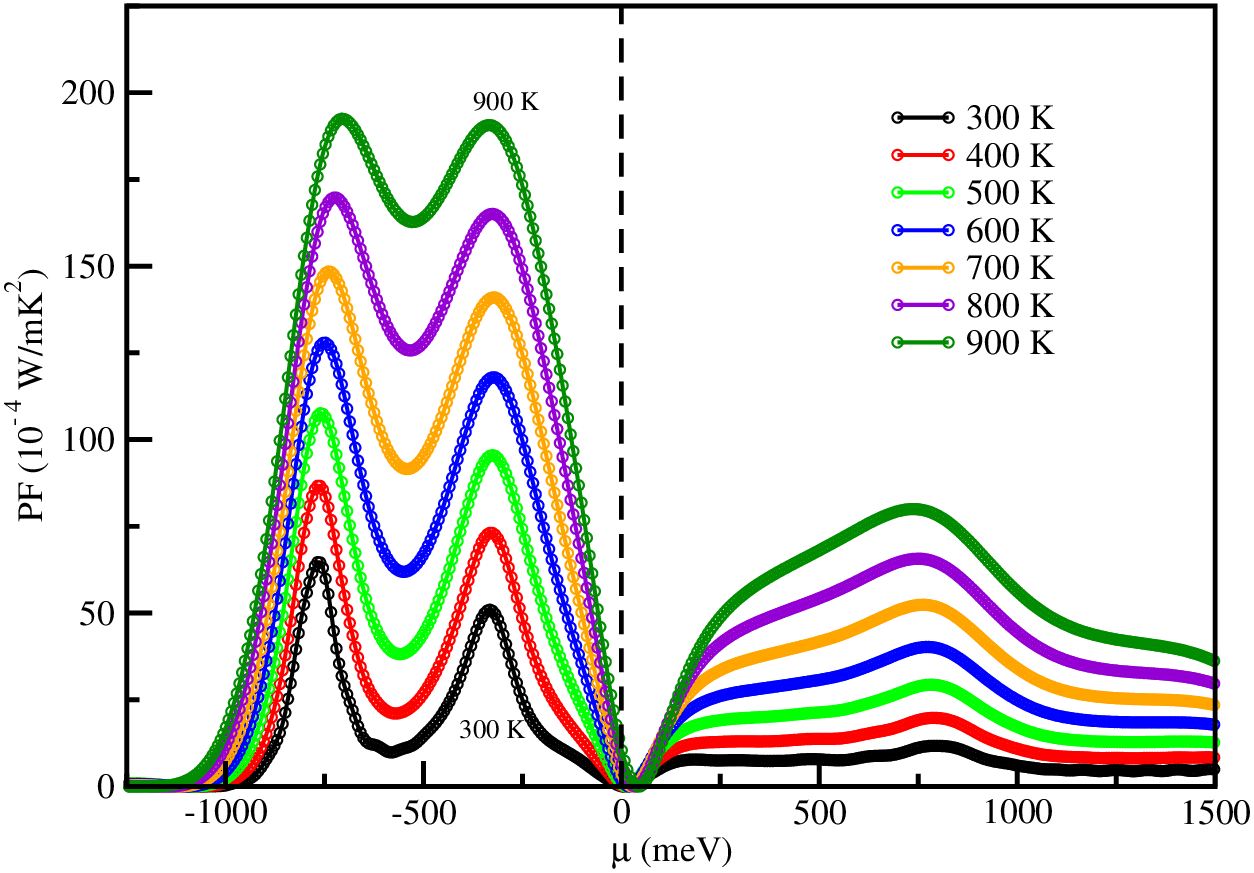} 
\caption{\label{Fig.8}\small{Variation of Power Factor with Chemical Potential}} 
\end{figure}
   
 



\section{Conclusions}
In summary, we have studied temperature dependent TE properties of PbSe by using both experimental and DFT technique in the high temperature range.
The measured value of \emph{S} at 300 K is $\sim$ 198 $\mu$V K$^{-1}$ and reaches to the value of 266 $\mu$V K$^{-1}$ at 500 K. The magnitude of \emph{S} increases in the whole temperature range, rate of increment in \emph{S} decreases after 460 K. The positive \emph{S} indicates the p-type behaviour of PbSe. The value of \emph{$\sigma$} increases linearly with increasing temperature. The observed values of \emph{$\sigma$} at 300 and 500 K is $\sim $ 0.35 $\times $ 10$^{3}$ and $\sim$ 0.58 $\times$ 10$^{3}$ $\Omega$$^{-1}$ m$^{-1}$, respectively. The \emph{$\kappa$} value increases from 300 to 460 K and then decreases upto 500 K. The value of \emph{$\kappa$} at 300, 460 and 500 K is $\sim$ 0.74, 1.10 and 1.07 W m$^{-1}$ K$^{-1}$, respectively. 
Additionaly, the TE properties have also been investigated using DFT and semi-classical Boltzmann transport theory. The electronic structure calculations are done with and without considering the SOC effect. Dispersion plot indicates that PbSe is a direct band gap semiconductor, with band gap values of 0.16 and 0.27 eV with SOC and without considering SOC, respectively. 
The calculated and experimental values of \emph{S} and \emph{$\sigma$} are found to be in good agreement.
Considering \emph{$\tau$} = 1.57 $\times$ 10$^{-16}$ seconds, we have evaluated the value of \emph{$\kappa{_e}$}. 
 The experimental \emph{$\kappa{_e}$} (using Wiedemann-Franz Law) shows that major contribution in total \emph{$\kappa$} is because of the lattice part. The experimental and calculated values of PF gives good match. The maximum observed value of PF has been found to be $\sim$ 192 $\times$ 10$^{-4}$ W/mK$^{2}$ at 900 K. This work propose that DFT based electronic structure calculations can be used in understanding the experimentally obtained TE properties of PbSe. Therefore, this approach can also be used to understand the transport behaviour of other chalcogenides.

\bibliography{ref}

\begin{thebibliography}{41}%
\makeatletter
\providecommand \@ifxundefined [1]{%
 \@ifx{#1\undefined}
}%
\providecommand \@ifnum [1]{%
 \ifnum #1\expandafter \@firstoftwo
 \else \expandafter \@secondoftwo
 \fi
}%
\providecommand \@ifx [1]{%
 \ifx #1\expandafter \@firstoftwo
 \else \expandafter \@secondoftwo
 \fi
}%
\providecommand \natexlab [1]{#1}%
\providecommand \enquote  [1]{``#1''}%
\providecommand \bibnamefont  [1]{#1}%
\providecommand \bibfnamefont [1]{#1}%
\providecommand \citenamefont [1]{#1}%
\providecommand \href@noop [0]{\@secondoftwo}%
\providecommand \href [0]{\begingroup \@sanitize@url \@href}%
\providecommand \@href[1]{\@@startlink{#1}\@@href}%
\providecommand \@@href[1]{\endgroup#1\@@endlink}%
\providecommand \@sanitize@url [0]{\catcode `\\12\catcode `\$12\catcode
  `\&12\catcode `\#12\catcode `\^12\catcode `\_12\catcode `\%12\relax}%
\providecommand \@@startlink[1]{}%
\providecommand \@@endlink[0]{}%
\providecommand \url  [0]{\begingroup\@sanitize@url \@url }%
\providecommand \@url [1]{\endgroup\@href {#1}{\urlprefix }}%
\providecommand \urlprefix  [0]{URL }%
\providecommand \Eprint [0]{\href }%
\providecommand \doibase [0]{http://dx.doi.org/}%
\providecommand \selectlanguage [0]{\@gobble}%
\providecommand \bibinfo  [0]{\@secondoftwo}%
\providecommand \bibfield  [0]{\@secondoftwo}%
\providecommand \translation [1]{[#1]}%
\providecommand \BibitemOpen [0]{}%
\providecommand \bibitemStop [0]{}%
\providecommand \bibitemNoStop [0]{.\EOS\space}%
\providecommand \EOS [0]{\spacefactor3000\relax}%
\providecommand \BibitemShut  [1]{\csname bibitem#1\endcsname}%
\let\auto@bib@innerbib\@empty
\bibitem [{\citenamefont {Mahan}\ and\ \citenamefont
  {Sofo}(1996)}]{mahan1996best}%
  \BibitemOpen
  \bibfield  {author} {\bibinfo {author} {\bibfnamefont {G.}~\bibnamefont
  {Mahan}}\ and\ \bibinfo {author} {\bibfnamefont {J.}~\bibnamefont {Sofo}},\
  }\href@noop {} {\bibfield  {journal} {\bibinfo  {journal} {Proceedings of the
  National Academy of Sciences}\ }\textbf {\bibinfo {volume} {93}},\ \bibinfo
  {pages} {7436} (\bibinfo {year} {1996})}\BibitemShut {NoStop}%
\bibitem [{\citenamefont {Pei}\ \emph {et~al.}(2011)\citenamefont {Pei},
  \citenamefont {Shi}, \citenamefont {LaLonde}, \citenamefont {Wang},
  \citenamefont {Chen},\ and\ \citenamefont {Snyder}}]{pei2011convergence}%
  \BibitemOpen
  \bibfield  {author} {\bibinfo {author} {\bibfnamefont {Y.}~\bibnamefont
  {Pei}}, \bibinfo {author} {\bibfnamefont {X.}~\bibnamefont {Shi}}, \bibinfo
  {author} {\bibfnamefont {A.}~\bibnamefont {LaLonde}}, \bibinfo {author}
  {\bibfnamefont {H.}~\bibnamefont {Wang}}, \bibinfo {author} {\bibfnamefont
  {L.}~\bibnamefont {Chen}}, \ and\ \bibinfo {author} {\bibfnamefont {G.~J.}\
  \bibnamefont {Snyder}},\ }\href@noop {} {\bibfield  {journal} {\bibinfo
  {journal} {Nature}\ }\textbf {\bibinfo {volume} {473}},\ \bibinfo {pages}
  {66} (\bibinfo {year} {2011})}\BibitemShut {NoStop}%
\bibitem [{\citenamefont {Pandey}\ and\ \citenamefont
  {Pandey}(2023)}]{pandey2023exploring}%
  \BibitemOpen
  \bibfield  {author} {\bibinfo {author} {\bibfnamefont {A.}~\bibnamefont
  {Pandey}}\ and\ \bibinfo {author} {\bibfnamefont {S.~K.}\ \bibnamefont
  {Pandey}},\ }\href@noop {} {\bibfield  {journal} {\bibinfo  {journal} {arXiv
  preprint arXiv:2312.10449}\ } (\bibinfo {year} {2023})}\BibitemShut {NoStop}%
\bibitem [{\citenamefont {Gayner}\ and\ \citenamefont
  {Kar}(2016)}]{gayner2016recent}%
  \BibitemOpen
  \bibfield  {author} {\bibinfo {author} {\bibfnamefont {C.}~\bibnamefont
  {Gayner}}\ and\ \bibinfo {author} {\bibfnamefont {K.~K.}\ \bibnamefont
  {Kar}},\ }\href@noop {} {\bibfield  {journal} {\bibinfo  {journal} {Progress
  in Materials Science}\ }\textbf {\bibinfo {volume} {83}},\ \bibinfo {pages}
  {330} (\bibinfo {year} {2016})}\BibitemShut {NoStop}%
\bibitem [{\citenamefont {Tan}\ \emph {et~al.}(2016{\natexlab{a}})\citenamefont
  {Tan}, \citenamefont {Zhao},\ and\ \citenamefont
  {Kanatzidis}}]{tan2016rationally}%
  \BibitemOpen
  \bibfield  {author} {\bibinfo {author} {\bibfnamefont {G.}~\bibnamefont
  {Tan}}, \bibinfo {author} {\bibfnamefont {L.-D.}\ \bibnamefont {Zhao}}, \
  and\ \bibinfo {author} {\bibfnamefont {M.~G.}\ \bibnamefont {Kanatzidis}},\
  }\href@noop {} {\bibfield  {journal} {\bibinfo  {journal} {Chemical reviews}\
  }\textbf {\bibinfo {volume} {116}},\ \bibinfo {pages} {12123} (\bibinfo
  {year} {2016}{\natexlab{a}})}\BibitemShut {NoStop}%
\bibitem [{\citenamefont {Qin}\ \emph {et~al.}(2021)\citenamefont {Qin},
  \citenamefont {Wang}, \citenamefont {Liu}, \citenamefont {Qin}, \citenamefont
  {Dong}, \citenamefont {Luo}, \citenamefont {Li}, \citenamefont {Liu},
  \citenamefont {Tan}, \citenamefont {Tang} \emph {et~al.}}]{qin2021power}%
  \BibitemOpen
  \bibfield  {author} {\bibinfo {author} {\bibfnamefont {B.}~\bibnamefont
  {Qin}}, \bibinfo {author} {\bibfnamefont {D.}~\bibnamefont {Wang}}, \bibinfo
  {author} {\bibfnamefont {X.}~\bibnamefont {Liu}}, \bibinfo {author}
  {\bibfnamefont {Y.}~\bibnamefont {Qin}}, \bibinfo {author} {\bibfnamefont
  {J.-F.}\ \bibnamefont {Dong}}, \bibinfo {author} {\bibfnamefont
  {J.}~\bibnamefont {Luo}}, \bibinfo {author} {\bibfnamefont {J.-W.}\
  \bibnamefont {Li}}, \bibinfo {author} {\bibfnamefont {W.}~\bibnamefont
  {Liu}}, \bibinfo {author} {\bibfnamefont {G.}~\bibnamefont {Tan}}, \bibinfo
  {author} {\bibfnamefont {X.}~\bibnamefont {Tang}},  \emph {et~al.},\
  }\href@noop {} {\bibfield  {journal} {\bibinfo  {journal} {Science}\ }\textbf
  {\bibinfo {volume} {373}},\ \bibinfo {pages} {556} (\bibinfo {year}
  {2021})}\BibitemShut {NoStop}%
\bibitem [{\citenamefont {Sun}\ \emph {et~al.}(2022)\citenamefont {Sun},
  \citenamefont {Zhang}, \citenamefont {Fan}, \citenamefont {Tang},\ and\
  \citenamefont {Tan}}]{sun2022strategies}%
  \BibitemOpen
  \bibfield  {author} {\bibinfo {author} {\bibfnamefont {J.}~\bibnamefont
  {Sun}}, \bibinfo {author} {\bibfnamefont {Y.}~\bibnamefont {Zhang}}, \bibinfo
  {author} {\bibfnamefont {Y.}~\bibnamefont {Fan}}, \bibinfo {author}
  {\bibfnamefont {X.}~\bibnamefont {Tang}}, \ and\ \bibinfo {author}
  {\bibfnamefont {G.}~\bibnamefont {Tan}},\ }\href@noop {} {\bibfield
  {journal} {\bibinfo  {journal} {Chemical Engineering Journal}\ }\textbf
  {\bibinfo {volume} {431}},\ \bibinfo {pages} {133699} (\bibinfo {year}
  {2022})}\BibitemShut {NoStop}%
\bibitem [{\citenamefont {Snyder}\ and\ \citenamefont
  {Toberer}(2008)}]{snyder2008complex}%
  \BibitemOpen
  \bibfield  {author} {\bibinfo {author} {\bibfnamefont {G.~J.}\ \bibnamefont
  {Snyder}}\ and\ \bibinfo {author} {\bibfnamefont {E.~S.}\ \bibnamefont
  {Toberer}},\ }\href@noop {} {\bibfield  {journal} {\bibinfo  {journal}
  {Nature materials}\ }\textbf {\bibinfo {volume} {7}},\ \bibinfo {pages} {105}
  (\bibinfo {year} {2008})}\BibitemShut {NoStop}%
\bibitem [{\citenamefont {Bano}\ \emph {et~al.}(2020)\citenamefont {Bano},
  \citenamefont {Kumar}, \citenamefont {Govind}, \citenamefont {Khan},
  \citenamefont {Ashok},\ and\ \citenamefont {Misra}}]{bano2020room}%
  \BibitemOpen
  \bibfield  {author} {\bibinfo {author} {\bibfnamefont {S.}~\bibnamefont
  {Bano}}, \bibinfo {author} {\bibfnamefont {A.}~\bibnamefont {Kumar}},
  \bibinfo {author} {\bibfnamefont {B.}~\bibnamefont {Govind}}, \bibinfo
  {author} {\bibfnamefont {A.~H.}\ \bibnamefont {Khan}}, \bibinfo {author}
  {\bibfnamefont {A.}~\bibnamefont {Ashok}}, \ and\ \bibinfo {author}
  {\bibfnamefont {D.}~\bibnamefont {Misra}},\ }\href@noop {} {\bibfield
  {journal} {\bibinfo  {journal} {Journal of Materials Science: Materials in
  Electronics}\ }\textbf {\bibinfo {volume} {31}},\ \bibinfo {pages} {8607}
  (\bibinfo {year} {2020})}\BibitemShut {NoStop}%
\bibitem [{\citenamefont {Zhu}\ \emph {et~al.}(2021)\citenamefont {Zhu},
  \citenamefont {Wang}, \citenamefont {Cui},\ and\ \citenamefont
  {He}}]{zhu2021point}%
  \BibitemOpen
  \bibfield  {author} {\bibinfo {author} {\bibfnamefont {B.}~\bibnamefont
  {Zhu}}, \bibinfo {author} {\bibfnamefont {W.}~\bibnamefont {Wang}}, \bibinfo
  {author} {\bibfnamefont {J.}~\bibnamefont {Cui}}, \ and\ \bibinfo {author}
  {\bibfnamefont {J.}~\bibnamefont {He}},\ }\href@noop {} {\bibfield  {journal}
  {\bibinfo  {journal} {Small}\ }\textbf {\bibinfo {volume} {17}},\ \bibinfo
  {pages} {2101328} (\bibinfo {year} {2021})}\BibitemShut {NoStop}%
\bibitem [{\citenamefont {Tan}\ \emph {et~al.}(2016{\natexlab{b}})\citenamefont
  {Tan}, \citenamefont {Shi}, \citenamefont {Hao}, \citenamefont {Zhao},
  \citenamefont {Chi}, \citenamefont {Zhang}, \citenamefont {Uher},
  \citenamefont {Wolverton}, \citenamefont {Dravid},\ and\ \citenamefont
  {Kanatzidis}}]{tan2016non}%
  \BibitemOpen
  \bibfield  {author} {\bibinfo {author} {\bibfnamefont {G.}~\bibnamefont
  {Tan}}, \bibinfo {author} {\bibfnamefont {F.}~\bibnamefont {Shi}}, \bibinfo
  {author} {\bibfnamefont {S.}~\bibnamefont {Hao}}, \bibinfo {author}
  {\bibfnamefont {L.-D.}\ \bibnamefont {Zhao}}, \bibinfo {author}
  {\bibfnamefont {H.}~\bibnamefont {Chi}}, \bibinfo {author} {\bibfnamefont
  {X.}~\bibnamefont {Zhang}}, \bibinfo {author} {\bibfnamefont
  {C.}~\bibnamefont {Uher}}, \bibinfo {author} {\bibfnamefont {C.}~\bibnamefont
  {Wolverton}}, \bibinfo {author} {\bibfnamefont {V.~P.}\ \bibnamefont
  {Dravid}}, \ and\ \bibinfo {author} {\bibfnamefont {M.~G.}\ \bibnamefont
  {Kanatzidis}},\ }\href@noop {} {\bibfield  {journal} {\bibinfo  {journal}
  {Nature communications}\ }\textbf {\bibinfo {volume} {7}},\ \bibinfo {pages}
  {12167} (\bibinfo {year} {2016}{\natexlab{b}})}\BibitemShut {NoStop}%
\bibitem [{\citenamefont {Liu}\ \emph {et~al.}(2022)\citenamefont {Liu},
  \citenamefont {Sun}, \citenamefont {Zhong}, \citenamefont {Deng},
  \citenamefont {Gan}, \citenamefont {Lv}, \citenamefont {Shi}, \citenamefont
  {Chen},\ and\ \citenamefont {Ang}}]{liu2022high}%
  \BibitemOpen
  \bibfield  {author} {\bibinfo {author} {\bibfnamefont {H.-T.}\ \bibnamefont
  {Liu}}, \bibinfo {author} {\bibfnamefont {Q.}~\bibnamefont {Sun}}, \bibinfo
  {author} {\bibfnamefont {Y.}~\bibnamefont {Zhong}}, \bibinfo {author}
  {\bibfnamefont {Q.}~\bibnamefont {Deng}}, \bibinfo {author} {\bibfnamefont
  {L.}~\bibnamefont {Gan}}, \bibinfo {author} {\bibfnamefont {F.-L.}\
  \bibnamefont {Lv}}, \bibinfo {author} {\bibfnamefont {X.-L.}\ \bibnamefont
  {Shi}}, \bibinfo {author} {\bibfnamefont {Z.-G.}\ \bibnamefont {Chen}}, \
  and\ \bibinfo {author} {\bibfnamefont {R.}~\bibnamefont {Ang}},\ }\href@noop
  {} {\bibfield  {journal} {\bibinfo  {journal} {Nano Energy}\ }\textbf
  {\bibinfo {volume} {91}},\ \bibinfo {pages} {106706} (\bibinfo {year}
  {2022})}\BibitemShut {NoStop}%
\bibitem [{\citenamefont {Khokhlov}(2021)}]{khokhlov2021lead}%
  \BibitemOpen
  \bibfield  {author} {\bibinfo {author} {\bibfnamefont {D.}~\bibnamefont
  {Khokhlov}},\ }\href@noop {} {\emph {\bibinfo {title} {Lead chalcogenides:
  physics and applications}}}\ (\bibinfo  {publisher} {Routledge},\ \bibinfo
  {year} {2021})\BibitemShut {NoStop}%
\bibitem [{\citenamefont {Ravich}(2013)}]{ravich2013semiconducting}%
  \BibitemOpen
  \bibfield  {author} {\bibinfo {author} {\bibfnamefont {I.~I.}\ \bibnamefont
  {Ravich}},\ }\href@noop {} {\emph {\bibinfo {title} {Semiconducting lead
  chalcogenides}}},\ Vol.~\bibinfo {volume} {5}\ (\bibinfo  {publisher}
  {Springer Science \& Business Media},\ \bibinfo {year} {2013})\BibitemShut
  {NoStop}%
\bibitem [{\citenamefont {Achimovi{\v{c}}ov{\'a}}\ \emph
  {et~al.}(2009)\citenamefont {Achimovi{\v{c}}ov{\'a}}, \citenamefont {Daneu},
  \citenamefont {Re{\v{c}}nik}, \citenamefont {{\v{D}}uri{\v{s}}in},
  \citenamefont {Peter}, \citenamefont {Fabi{\'a}n}, \citenamefont
  {Kov{\'a}{\v{c}}},\ and\ \citenamefont
  {{\v{S}}atka}}]{achimovivcova2009characterization}%
  \BibitemOpen
  \bibfield  {author} {\bibinfo {author} {\bibfnamefont {M.}~\bibnamefont
  {Achimovi{\v{c}}ov{\'a}}}, \bibinfo {author} {\bibfnamefont {N.}~\bibnamefont
  {Daneu}}, \bibinfo {author} {\bibfnamefont {A.}~\bibnamefont {Re{\v{c}}nik}},
  \bibinfo {author} {\bibfnamefont {J.}~\bibnamefont {{\v{D}}uri{\v{s}}in}},
  \bibinfo {author} {\bibfnamefont {B.}~\bibnamefont {Peter}}, \bibinfo
  {author} {\bibfnamefont {M.}~\bibnamefont {Fabi{\'a}n}}, \bibinfo {author}
  {\bibfnamefont {J.}~\bibnamefont {Kov{\'a}{\v{c}}}}, \ and\ \bibinfo {author}
  {\bibfnamefont {A.}~\bibnamefont {{\v{S}}atka}},\ }\href@noop {} {\bibfield
  {journal} {\bibinfo  {journal} {Chemical Papers}\ }\textbf {\bibinfo {volume}
  {63}},\ \bibinfo {pages} {562} (\bibinfo {year} {2009})}\BibitemShut
  {NoStop}%
\bibitem [{\citenamefont {Alekseeva}\ \emph {et~al.}(1996)\citenamefont
  {Alekseeva}, \citenamefont {Gurieva}, \citenamefont {Konstantinov},
  \citenamefont {Prokof'eva},\ and\ \citenamefont
  {Fedorov}}]{alekseeva1996thermoelectric}%
  \BibitemOpen
  \bibfield  {author} {\bibinfo {author} {\bibfnamefont {G.}~\bibnamefont
  {Alekseeva}}, \bibinfo {author} {\bibfnamefont {E.}~\bibnamefont {Gurieva}},
  \bibinfo {author} {\bibfnamefont {P.}~\bibnamefont {Konstantinov}}, \bibinfo
  {author} {\bibfnamefont {L.}~\bibnamefont {Prokof'eva}}, \ and\ \bibinfo
  {author} {\bibfnamefont {M.}~\bibnamefont {Fedorov}},\ }\href@noop {}
  {\bibfield  {journal} {\bibinfo  {journal} {Semiconductors}\ }\textbf
  {\bibinfo {volume} {30}},\ \bibinfo {pages} {1125} (\bibinfo {year}
  {1996})}\BibitemShut {NoStop}%
\bibitem [{\citenamefont {Ibrahim}\ \emph {et~al.}(2008)\citenamefont
  {Ibrahim}, \citenamefont {Saleh}, \citenamefont {Ibrahim},\ and\
  \citenamefont {Hakeem}}]{ibrahim2008electrical}%
  \BibitemOpen
  \bibfield  {author} {\bibinfo {author} {\bibfnamefont {M.}~\bibnamefont
  {Ibrahim}}, \bibinfo {author} {\bibfnamefont {S.}~\bibnamefont {Saleh}},
  \bibinfo {author} {\bibfnamefont {E.}~\bibnamefont {Ibrahim}}, \ and\
  \bibinfo {author} {\bibfnamefont {A.~A.}\ \bibnamefont {Hakeem}},\
  }\href@noop {} {\bibfield  {journal} {\bibinfo  {journal} {Journal of alloys
  and compounds}\ }\textbf {\bibinfo {volume} {452}},\ \bibinfo {pages} {200}
  (\bibinfo {year} {2008})}\BibitemShut {NoStop}%
\bibitem [{\citenamefont {Du}\ \emph {et~al.}(2020)\citenamefont {Du},
  \citenamefont {Su}, \citenamefont {Li}, \citenamefont {Li}, \citenamefont
  {Hu}, \citenamefont {Fan}, \citenamefont {Hu},\ and\ \citenamefont
  {Zhu}}]{du2020rapid}%
  \BibitemOpen
  \bibfield  {author} {\bibinfo {author} {\bibfnamefont {J.}~\bibnamefont
  {Du}}, \bibinfo {author} {\bibfnamefont {T.}~\bibnamefont {Su}}, \bibinfo
  {author} {\bibfnamefont {H.}~\bibnamefont {Li}}, \bibinfo {author}
  {\bibfnamefont {S.}~\bibnamefont {Li}}, \bibinfo {author} {\bibfnamefont
  {M.}~\bibnamefont {Hu}}, \bibinfo {author} {\bibfnamefont {H.}~\bibnamefont
  {Fan}}, \bibinfo {author} {\bibfnamefont {Q.}~\bibnamefont {Hu}}, \ and\
  \bibinfo {author} {\bibfnamefont {H.}~\bibnamefont {Zhu}},\ }\href@noop {}
  {\bibfield  {journal} {\bibinfo  {journal} {Journal of Materials Science:
  Materials in Electronics}\ }\textbf {\bibinfo {volume} {31}},\ \bibinfo
  {pages} {6855} (\bibinfo {year} {2020})}\BibitemShut {NoStop}%
\bibitem [{\citenamefont {Chen}\ \emph {et~al.}(2018)\citenamefont {Chen},
  \citenamefont {Li},\ and\ \citenamefont {Sun}}]{chen2018effects}%
  \BibitemOpen
  \bibfield  {author} {\bibinfo {author} {\bibfnamefont {B.}~\bibnamefont
  {Chen}}, \bibinfo {author} {\bibfnamefont {Y.}~\bibnamefont {Li}}, \ and\
  \bibinfo {author} {\bibfnamefont {Z.-Y.}\ \bibnamefont {Sun}},\ }\href@noop
  {} {\bibfield  {journal} {\bibinfo  {journal} {Journal of Electronic
  Materials}\ }\textbf {\bibinfo {volume} {47}},\ \bibinfo {pages} {3099}
  (\bibinfo {year} {2018})}\BibitemShut {NoStop}%
\bibitem [{\citenamefont {Wang}\ \emph {et~al.}(2011)\citenamefont {Wang},
  \citenamefont {Zheng}, \citenamefont {Luo}, \citenamefont {She},
  \citenamefont {Li},\ and\ \citenamefont {Tang}}]{wang2011exploring}%
  \BibitemOpen
  \bibfield  {author} {\bibinfo {author} {\bibfnamefont {S.}~\bibnamefont
  {Wang}}, \bibinfo {author} {\bibfnamefont {G.}~\bibnamefont {Zheng}},
  \bibinfo {author} {\bibfnamefont {T.}~\bibnamefont {Luo}}, \bibinfo {author}
  {\bibfnamefont {X.}~\bibnamefont {She}}, \bibinfo {author} {\bibfnamefont
  {H.}~\bibnamefont {Li}}, \ and\ \bibinfo {author} {\bibfnamefont
  {X.}~\bibnamefont {Tang}},\ }\href@noop {} {\bibfield  {journal} {\bibinfo
  {journal} {Journal of Physics D: Applied Physics}\ }\textbf {\bibinfo
  {volume} {44}},\ \bibinfo {pages} {475304} (\bibinfo {year}
  {2011})}\BibitemShut {NoStop}%
\bibitem [{\citenamefont {Tang}\ \emph {et~al.}(2022)\citenamefont {Tang},
  \citenamefont {Bai}, \citenamefont {Wu}, \citenamefont {Luo}, \citenamefont
  {Wang}, \citenamefont {Yang},\ and\ \citenamefont
  {Zhao}}]{tang2022honeycomb}%
  \BibitemOpen
  \bibfield  {author} {\bibinfo {author} {\bibfnamefont {S.}~\bibnamefont
  {Tang}}, \bibinfo {author} {\bibfnamefont {S.}~\bibnamefont {Bai}}, \bibinfo
  {author} {\bibfnamefont {M.}~\bibnamefont {Wu}}, \bibinfo {author}
  {\bibfnamefont {D.}~\bibnamefont {Luo}}, \bibinfo {author} {\bibfnamefont
  {D.}~\bibnamefont {Wang}}, \bibinfo {author} {\bibfnamefont {S.}~\bibnamefont
  {Yang}}, \ and\ \bibinfo {author} {\bibfnamefont {L.-D.}\ \bibnamefont
  {Zhao}},\ }\href@noop {} {\bibfield  {journal} {\bibinfo  {journal}
  {Materials Today Energy}\ }\textbf {\bibinfo {volume} {23}},\ \bibinfo
  {pages} {100914} (\bibinfo {year} {2022})}\BibitemShut {NoStop}%
\bibitem [{\citenamefont {Parker}\ and\ \citenamefont
  {Singh}(2010)}]{parker2010high}%
  \BibitemOpen
  \bibfield  {author} {\bibinfo {author} {\bibfnamefont {D.}~\bibnamefont
  {Parker}}\ and\ \bibinfo {author} {\bibfnamefont {D.~J.}\ \bibnamefont
  {Singh}},\ }\href@noop {} {\bibfield  {journal} {\bibinfo  {journal}
  {Physical Review B}\ }\textbf {\bibinfo {volume} {82}},\ \bibinfo {pages}
  {035204} (\bibinfo {year} {2010})}\BibitemShut {NoStop}%
\bibitem [{\citenamefont {Gayner}\ \emph {et~al.}(2016)\citenamefont {Gayner},
  \citenamefont {Sharma}, \citenamefont {Das},\ and\ \citenamefont
  {Kar}}]{gayner2016boost}%
  \BibitemOpen
  \bibfield  {author} {\bibinfo {author} {\bibfnamefont {C.}~\bibnamefont
  {Gayner}}, \bibinfo {author} {\bibfnamefont {R.}~\bibnamefont {Sharma}},
  \bibinfo {author} {\bibfnamefont {M.~K.}\ \bibnamefont {Das}}, \ and\
  \bibinfo {author} {\bibfnamefont {K.~K.}\ \bibnamefont {Kar}},\ }\href@noop
  {} {\bibfield  {journal} {\bibinfo  {journal} {Journal of Applied Physics}\
  }\textbf {\bibinfo {volume} {120}} (\bibinfo {year} {2016})}\BibitemShut
  {NoStop}%
\bibitem [{\citenamefont {Khan}\ \emph {et~al.}(2016)\citenamefont {Khan},
  \citenamefont {Khan}, \citenamefont {Ahmad},\ and\ \citenamefont
  {Ali}}]{khan2016thermoelectric}%
  \BibitemOpen
  \bibfield  {author} {\bibinfo {author} {\bibfnamefont {A.~A.}\ \bibnamefont
  {Khan}}, \bibinfo {author} {\bibfnamefont {I.}~\bibnamefont {Khan}}, \bibinfo
  {author} {\bibfnamefont {I.}~\bibnamefont {Ahmad}}, \ and\ \bibinfo {author}
  {\bibfnamefont {Z.}~\bibnamefont {Ali}},\ }\href@noop {} {\bibfield
  {journal} {\bibinfo  {journal} {Materials Science in Semiconductor
  Processing}\ }\textbf {\bibinfo {volume} {48}},\ \bibinfo {pages} {85}
  (\bibinfo {year} {2016})}\BibitemShut {NoStop}%
\bibitem [{\citenamefont {Sk}\ \emph {et~al.}(2022)\citenamefont {Sk},
  \citenamefont {Pandey},\ and\ \citenamefont {Pandey}}]{sk2022instrument}%
  \BibitemOpen
  \bibfield  {author} {\bibinfo {author} {\bibfnamefont {S.}~\bibnamefont
  {Sk}}, \bibinfo {author} {\bibfnamefont {A.}~\bibnamefont {Pandey}}, \ and\
  \bibinfo {author} {\bibfnamefont {S.~K.}\ \bibnamefont {Pandey}},\
  }\href@noop {} {\bibfield  {journal} {\bibinfo  {journal} {Review of
  Scientific Instruments}\ }\textbf {\bibinfo {volume} {93}},\ \bibinfo {pages}
  {043902} (\bibinfo {year} {2022})}\BibitemShut {NoStop}%
\bibitem [{\citenamefont {Singh}\ and\ \citenamefont
  {Pandey}(2018)}]{singh2018fabrication}%
  \BibitemOpen
  \bibfield  {author} {\bibinfo {author} {\bibfnamefont {S.}~\bibnamefont
  {Singh}}\ and\ \bibinfo {author} {\bibfnamefont {S.~K.}\ \bibnamefont
  {Pandey}},\ }\href@noop {} {\bibfield  {journal} {\bibinfo  {journal} {IEEE
  Transactions on Instrumentation and Measurement}\ }\textbf {\bibinfo {volume}
  {67}},\ \bibinfo {pages} {2169} (\bibinfo {year} {2018})}\BibitemShut
  {NoStop}%
\bibitem [{\citenamefont {Duwez}\ \emph {et~al.}(1960)\citenamefont {Duwez},
  \citenamefont {Willens},\ and\ \citenamefont
  {Klement~Jr}}]{duwez1960continuous}%
  \BibitemOpen
  \bibfield  {author} {\bibinfo {author} {\bibfnamefont {P.}~\bibnamefont
  {Duwez}}, \bibinfo {author} {\bibfnamefont {R.}~\bibnamefont {Willens}}, \
  and\ \bibinfo {author} {\bibfnamefont {W.}~\bibnamefont {Klement~Jr}},\
  }\href@noop {} {\bibfield  {journal} {\bibinfo  {journal} {Journal of Applied
  Physics}\ }\textbf {\bibinfo {volume} {31}},\ \bibinfo {pages} {1136}
  (\bibinfo {year} {1960})}\BibitemShut {NoStop}%
\bibitem [{\citenamefont {Anantharaman}\ and\ \citenamefont
  {Suryanarayana}(1971)}]{anantharaman1971decade}%
  \BibitemOpen
  \bibfield  {author} {\bibinfo {author} {\bibfnamefont {T.}~\bibnamefont
  {Anantharaman}}\ and\ \bibinfo {author} {\bibfnamefont {C.}~\bibnamefont
  {Suryanarayana}},\ }\href@noop {} {\bibfield  {journal} {\bibinfo  {journal}
  {Journal of Materials Science}\ }\textbf {\bibinfo {volume} {6}},\ \bibinfo
  {pages} {1111} (\bibinfo {year} {1971})}\BibitemShut {NoStop}%
\bibitem [{\citenamefont {Degen}\ \emph {et~al.}(2014)\citenamefont {Degen},
  \citenamefont {Sadki}, \citenamefont {Bron}, \citenamefont {K{\"o}nig},\ and\
  \citenamefont {N{\'e}nert}}]{degen2014highscore}%
  \BibitemOpen
  \bibfield  {author} {\bibinfo {author} {\bibfnamefont {T.}~\bibnamefont
  {Degen}}, \bibinfo {author} {\bibfnamefont {M.}~\bibnamefont {Sadki}},
  \bibinfo {author} {\bibfnamefont {E.}~\bibnamefont {Bron}}, \bibinfo {author}
  {\bibfnamefont {U.}~\bibnamefont {K{\"o}nig}}, \ and\ \bibinfo {author}
  {\bibfnamefont {G.}~\bibnamefont {N{\'e}nert}},\ }\href@noop {} {\bibfield
  {journal} {\bibinfo  {journal} {Powder diffraction}\ }\textbf {\bibinfo
  {volume} {29}},\ \bibinfo {pages} {S13} (\bibinfo {year} {2014})}\BibitemShut
  {NoStop}%
\bibitem [{\citenamefont {Bl{\"o}chl}(1994)}]{blochl1994projector}%
  \BibitemOpen
  \bibfield  {author} {\bibinfo {author} {\bibfnamefont {P.~E.}\ \bibnamefont
  {Bl{\"o}chl}},\ }\href@noop {} {\bibfield  {journal} {\bibinfo  {journal}
  {Physical review B}\ }\textbf {\bibinfo {volume} {50}},\ \bibinfo {pages}
  {17953} (\bibinfo {year} {1994})}\BibitemShut {NoStop}%
\bibitem [{\citenamefont {Gonze}\ \emph {et~al.}(2002)\citenamefont {Gonze},
  \citenamefont {Beuken}, \citenamefont {Caracas}, \citenamefont {Detraux},
  \citenamefont {Fuchs}, \citenamefont {Rignanese}, \citenamefont {Sindic},
  \citenamefont {Verstraete}, \citenamefont {Zerah}, \citenamefont {Jollet}
  \emph {et~al.}}]{gonze2002first}%
  \BibitemOpen
  \bibfield  {author} {\bibinfo {author} {\bibfnamefont {X.}~\bibnamefont
  {Gonze}}, \bibinfo {author} {\bibfnamefont {J.-M.}\ \bibnamefont {Beuken}},
  \bibinfo {author} {\bibfnamefont {R.}~\bibnamefont {Caracas}}, \bibinfo
  {author} {\bibfnamefont {F.}~\bibnamefont {Detraux}}, \bibinfo {author}
  {\bibfnamefont {M.}~\bibnamefont {Fuchs}}, \bibinfo {author} {\bibfnamefont
  {G.-M.}\ \bibnamefont {Rignanese}}, \bibinfo {author} {\bibfnamefont
  {L.}~\bibnamefont {Sindic}}, \bibinfo {author} {\bibfnamefont
  {M.}~\bibnamefont {Verstraete}}, \bibinfo {author} {\bibfnamefont
  {G.}~\bibnamefont {Zerah}}, \bibinfo {author} {\bibfnamefont
  {F.}~\bibnamefont {Jollet}},  \emph {et~al.},\ }\href@noop {} {\bibfield
  {journal} {\bibinfo  {journal} {Computational Materials Science}\ }\textbf
  {\bibinfo {volume} {25}},\ \bibinfo {pages} {478} (\bibinfo {year}
  {2002})}\BibitemShut {NoStop}%
\bibitem [{\citenamefont {Perdew}\ \emph {et~al.}(1996)\citenamefont {Perdew},
  \citenamefont {Burke},\ and\ \citenamefont
  {Ernzerhof}}]{perdew1996generalized}%
  \BibitemOpen
  \bibfield  {author} {\bibinfo {author} {\bibfnamefont {J.~P.}\ \bibnamefont
  {Perdew}}, \bibinfo {author} {\bibfnamefont {K.}~\bibnamefont {Burke}}, \
  and\ \bibinfo {author} {\bibfnamefont {M.}~\bibnamefont {Ernzerhof}},\
  }\href@noop {} {\bibfield  {journal} {\bibinfo  {journal} {Physical review
  letters}\ }\textbf {\bibinfo {volume} {77}},\ \bibinfo {pages} {3865}
  (\bibinfo {year} {1996})}\BibitemShut {NoStop}%
\bibitem [{\citenamefont {Madsen}\ and\ \citenamefont
  {Singh}(2006)}]{madsen2006boltztrap}%
  \BibitemOpen
  \bibfield  {author} {\bibinfo {author} {\bibfnamefont {G.~K.}\ \bibnamefont
  {Madsen}}\ and\ \bibinfo {author} {\bibfnamefont {D.~J.}\ \bibnamefont
  {Singh}},\ }\href@noop {} {\bibfield  {journal} {\bibinfo  {journal}
  {Computer Physics Communications}\ }\textbf {\bibinfo {volume} {175}},\
  \bibinfo {pages} {67} (\bibinfo {year} {2006})}\BibitemShut {NoStop}%
\bibitem [{\citenamefont {Nikolic}(1969)}]{nikolic1969solid}%
  \BibitemOpen
  \bibfield  {author} {\bibinfo {author} {\bibfnamefont {P.}~\bibnamefont
  {Nikolic}},\ }\href@noop {} {\bibfield  {journal} {\bibinfo  {journal}
  {Journal of Physics D: Applied Physics}\ }\textbf {\bibinfo {volume} {2}},\
  \bibinfo {pages} {383} (\bibinfo {year} {1969})}\BibitemShut {NoStop}%
\bibitem [{\citenamefont {Fan}\ \emph {et~al.}(2014)\citenamefont {Fan},
  \citenamefont {Su}, \citenamefont {Li}, \citenamefont {Zheng}, \citenamefont
  {Li}, \citenamefont {Hu}, \citenamefont {Zhou}, \citenamefont {Ma},\ and\
  \citenamefont {Jia}}]{fan2014high}%
  \BibitemOpen
  \bibfield  {author} {\bibinfo {author} {\bibfnamefont {H.}~\bibnamefont
  {Fan}}, \bibinfo {author} {\bibfnamefont {T.}~\bibnamefont {Su}}, \bibinfo
  {author} {\bibfnamefont {H.}~\bibnamefont {Li}}, \bibinfo {author}
  {\bibfnamefont {Y.}~\bibnamefont {Zheng}}, \bibinfo {author} {\bibfnamefont
  {S.}~\bibnamefont {Li}}, \bibinfo {author} {\bibfnamefont {M.}~\bibnamefont
  {Hu}}, \bibinfo {author} {\bibfnamefont {Y.}~\bibnamefont {Zhou}}, \bibinfo
  {author} {\bibfnamefont {H.}~\bibnamefont {Ma}}, \ and\ \bibinfo {author}
  {\bibfnamefont {X.}~\bibnamefont {Jia}},\ }\href@noop {} {\bibfield
  {journal} {\bibinfo  {journal} {Solid State Communications}\ }\textbf
  {\bibinfo {volume} {186}},\ \bibinfo {pages} {8} (\bibinfo {year}
  {2014})}\BibitemShut {NoStop}%
\bibitem [{\citenamefont {Francl}\ and\ \citenamefont
  {Kingery}(1954)}]{francl1954thermal}%
  \BibitemOpen
  \bibfield  {author} {\bibinfo {author} {\bibfnamefont {J.}~\bibnamefont
  {Francl}}\ and\ \bibinfo {author} {\bibfnamefont {W.}~\bibnamefont
  {Kingery}},\ }\href@noop {} {\bibfield  {journal} {\bibinfo  {journal}
  {Journal of the American ceramic Society}\ }\textbf {\bibinfo {volume}
  {37}},\ \bibinfo {pages} {99} (\bibinfo {year} {1954})}\BibitemShut {NoStop}%
\bibitem [{\citenamefont {Albanesi}\ \emph {et~al.}(2000)\citenamefont
  {Albanesi}, \citenamefont {Okoye}, \citenamefont {Rodr{\'\i}guez},
  \citenamefont {y~Blanca},\ and\ \citenamefont
  {Petukhov}}]{albanesi2000electronic}%
  \BibitemOpen
  \bibfield  {author} {\bibinfo {author} {\bibfnamefont {E.~A.}\ \bibnamefont
  {Albanesi}}, \bibinfo {author} {\bibfnamefont {C.}~\bibnamefont {Okoye}},
  \bibinfo {author} {\bibfnamefont {C.~O.}\ \bibnamefont {Rodr{\'\i}guez}},
  \bibinfo {author} {\bibfnamefont {E.~P.}\ \bibnamefont {y~Blanca}}, \ and\
  \bibinfo {author} {\bibfnamefont {A.}~\bibnamefont {Petukhov}},\ }\href@noop
  {} {\bibfield  {journal} {\bibinfo  {journal} {Physical Review B}\ }\textbf
  {\bibinfo {volume} {61}},\ \bibinfo {pages} {16589} (\bibinfo {year}
  {2000})}\BibitemShut {NoStop}%
\bibitem [{\citenamefont {Ashcroft}\ and\ \citenamefont
  {Mermin}(2011)}]{ashcroft2011solid}%
  \BibitemOpen
  \bibfield  {author} {\bibinfo {author} {\bibfnamefont {N.~W.}\ \bibnamefont
  {Ashcroft}}\ and\ \bibinfo {author} {\bibfnamefont {N.}~\bibnamefont
  {Mermin}},\ }\href@noop {} {\bibfield  {journal} {\bibinfo  {journal}
  {Cengage Learning: Boston, MA, USA}\ } (\bibinfo {year} {2011})}\BibitemShut
  {NoStop}%
\bibitem [{\citenamefont {Durczewski}\ and\ \citenamefont
  {Ausloos}(2000)}]{durczewski2000nontrivial}%
  \BibitemOpen
  \bibfield  {author} {\bibinfo {author} {\bibfnamefont {K.}~\bibnamefont
  {Durczewski}}\ and\ \bibinfo {author} {\bibfnamefont {M.}~\bibnamefont
  {Ausloos}},\ }\href@noop {} {\bibfield  {journal} {\bibinfo  {journal}
  {Physical Review B}\ }\textbf {\bibinfo {volume} {61}},\ \bibinfo {pages}
  {5303} (\bibinfo {year} {2000})}\BibitemShut {NoStop}%
\bibitem [{\citenamefont {Ahmad}\ and\ \citenamefont
  {Mahanti}(2010)}]{ahmad2010energy}%
  \BibitemOpen
  \bibfield  {author} {\bibinfo {author} {\bibfnamefont {S.}~\bibnamefont
  {Ahmad}}\ and\ \bibinfo {author} {\bibfnamefont {S.}~\bibnamefont
  {Mahanti}},\ }\href@noop {} {\bibfield  {journal} {\bibinfo  {journal}
  {Physical Review B}\ }\textbf {\bibinfo {volume} {81}},\ \bibinfo {pages}
  {165203} (\bibinfo {year} {2010})}\BibitemShut {NoStop}%
\bibitem [{\citenamefont {Zhang}\ \emph {et~al.}(2012)\citenamefont {Zhang},
  \citenamefont {Cao}, \citenamefont {Lukas}, \citenamefont {Liu},
  \citenamefont {Esfarjani}, \citenamefont {Opeil}, \citenamefont {Broido},
  \citenamefont {Parker}, \citenamefont {Singh}, \citenamefont {Chen} \emph
  {et~al.}}]{zhang2012study}%
  \BibitemOpen
  \bibfield  {author} {\bibinfo {author} {\bibfnamefont {Q.}~\bibnamefont
  {Zhang}}, \bibinfo {author} {\bibfnamefont {F.}~\bibnamefont {Cao}}, \bibinfo
  {author} {\bibfnamefont {K.}~\bibnamefont {Lukas}}, \bibinfo {author}
  {\bibfnamefont {W.}~\bibnamefont {Liu}}, \bibinfo {author} {\bibfnamefont
  {K.}~\bibnamefont {Esfarjani}}, \bibinfo {author} {\bibfnamefont
  {C.}~\bibnamefont {Opeil}}, \bibinfo {author} {\bibfnamefont
  {D.}~\bibnamefont {Broido}}, \bibinfo {author} {\bibfnamefont
  {D.}~\bibnamefont {Parker}}, \bibinfo {author} {\bibfnamefont {D.~J.}\
  \bibnamefont {Singh}}, \bibinfo {author} {\bibfnamefont {G.}~\bibnamefont
  {Chen}},  \emph {et~al.},\ }\href@noop {} {\bibfield  {journal} {\bibinfo
  {journal} {Journal of the American Chemical Society}\ }\textbf {\bibinfo
  {volume} {134}},\ \bibinfo {pages} {17731} (\bibinfo {year}
  {2012})}\BibitemShut {NoStop}%
\end{thebibliography}%
\bibliographystyle{apsrev4-1}

\end{document}